\documentclass[10pt, a4paper]{article}

\usepackage[final]{lrec2026} 
\usepackage{amsmath}
\usepackage{amssymb}
\usepackage{multirow}
\usepackage{booktabs}
\usepackage[table]{xcolor}
\usepackage{tabularx}
\usepackage{enumitem}
\usepackage{caption}

\title{ExplainableGuard: Interpretable Adversarial Defense for LLMs Using Chain-of-Thought Reasoning}

\name{Shaowei GUAN\textsuperscript{1}, Yu Zhai\textsuperscript{2}, Zhengyu ZHANG\textsuperscript{3}, Yanze Wang\textsuperscript{3}, Hin Chi Kwok\textsuperscript{1}} 

\address{ \textsuperscript{1}Centre for Smart Health, School of Nursing, The Hong Kong Polytechnic University, Hong Kong, China \\
         \textsuperscript{2}Department of Language Science and Technology, The Hong Kong Polytechnic University, Hong Kong, China\\
         \textsuperscript{3}Department of Electrical and Electronic Engineering, The Hong Kong Polytechnic University, Hong Kong, China \\
         \textsuperscript{1}spiderman.guan@connect.polyu.hk\\
         }

\abstract{
Large Language Models (LLMs) are increasingly vulnerable to adversarial attacks that can subtly manipulate their outputs. While various defense mechanisms have been proposed, many operate as black boxes, lacking transparency in their decision-making. This paper introduces ExplainableGuard, an interpretable adversarial defense framework leveraging the chain-of-thought (CoT) reasoning capabilities of DeepSeek-Reasoner. Our approach not only detects and neutralizes adversarial perturbations in text but also provides step-by-step explanations for each defense action. We demonstrate how tailored CoT prompts guide the LLM to perform a multi-faceted analysis (character, word, structural, semantic) and generate a purified output along with a human-readable justification. Preliminary results on GLUE Benchmark and IMDB Movie Reviews dataset show promising defense efficacy. Additionally, a human evaluation study reveals that ExplainableGuard’s explanations outperform ablated variants in clarity, specificity, and actionability, with a 72.5\% deployability-trust rating, underscoring its potential for more trustworthy LLM deployments.
 \\ \newline \Keywords{Chain-of-Thought, Explainable AI, Large Language Models, AI Security, Adversarial Defend} }

\begin{document}

\maketitleabstract

\section{Introduction}
Large Language Models (LLMs) such as GPT-4 \cite{achiam2023gpt}, Llama \citep{touvron2023llama}, and others have demonstrated remarkable capabilities across diverse natural language processing (NLP) tasks. However, their widespread adoption is hampered by their susceptibility to adversarial attacks \citep{Goodfellow2014ExplainingAH, jin2020bert}. These attacks involve crafting subtle, often human-imperceptible perturbations to input text, causing LLMs to produce erroneous, biased, or harmful outputs.

Existing defense strategies range from input pre-processing and adversarial training to detection mechanisms \citep{jia2019certified, zhu2019freelb}. While effective to some extent, many of these methods lack transparency. Understanding “why” a specific input was flagged or modified is crucial for building trust, debugging models, and iterating on security measures. This is particularly important in high-stakes applications.

To address this gap, we propose \textbf{ExplainableGuard}, a novel defense mechanism that utilizes the advanced reasoning abilities of a powerful LLM, DeepSeek-Reasoner, to not only defend against adversarial attacks but also to explain its defense process. Our core contribution lies in designing a structured, Chain-of-Thought (CoT) \citep{Wei2022ChainOT} prompting strategy that elicits detailed reasoning from the defense LLM. This reasoning breaks down the analysis into character, word, structural, and semantic levels, culminating in a decision, a purified text, and a comprehensive explanation.

The structure of this paper is as follows: Section \ref{sec:related_work} provides a review of prior research on adversarial attacks and defense. Section \ref{sec:methodology} describes the components of the ExplainableGuard architecture, details the Chain-of-Thought (CoT) prompting approach, and highlights its potential for robust and interpretable defense mechanisms. Section \ref{sec:experiments} outlines the evaluation design, while Section \ref{sec:results} Results presents both the quantitative and qualitative findings. The Limitations of the proposed approach are discussed in Section \ref{sec:dicussion}, and the paper concludes with final remarks in Section \ref{sec:conclusion}. This work aims to demonstrate that the proposed system effectively mitigates various types of adversarial attacks while providing valuable insights into its operational logic.

\section{Related Work}
\label{sec:related_work}

\subsection{Adversarial Attacks on LLMs}
Adversarial attacks on LLMs involve subtle perturbations of the input text that alter the predictions of the model without affecting the human-perceived meaning of the text \citep{xu2023llmfoolitself}. Adversarial attacks can be categorized into several levels. Character-level attacks involve manipulations such as typos, homoglyphs, or invisible characters \citep{Ebrahimi2018HotFlipWA}. Word-level attacks involve replacing words with synonyms, paraphrasing sentences, or inserting/deleting words \citep{Alzantot2018GeneratingNA}. More sophisticated attacks target sentence structure or semantics, including prompt injection and jailbreak techniques \citep{zou2023universaltransferableadversarialattacks}. These types of attacks expose the vulnerability of current LLMs, highlighting the need for more robust defenses \citep{wang2023adversarialattacksdefensesmachine}.

\subsection{Adversarial Defense Mechanisms}

Defense usually involves input cleaning (e.g., filtering special characters, spell checking), adversarial training (fine-tuning models of adversarial examples)\citep{goyal2023asurveyofadversarialdefenses, jia2019certified}. In the field of image classification, adversarial defense also involves authentication defense that provides robustness guarantees \citep{croce2020robustbench}. Some methods employ detector models to flag malicious inputs \citep{Mozes2023UseOI}. However, many of these defenses do not clearly explain why an input is considered adversarial or how an attack is eliminated. A similar work by Lin et al. \citep{lin2025largelanguagemodelsentinel} uses an LLM agent for adversarial purification and achieves satisfying performance. However, it lacks explainability, which represents a significant gap.

\subsection{Explainable AI (XAI) in NLP}

Explainability in natural language processing (NLP) seeks to render model predictions and internal reasoning processes transparent and interpretable \citep{Danilevsky2020ASO}. Traditional approaches include attention visualization, such as AttentionViz \citep{yeh2023attentionvizglobalviewtransformer}, and feature‐attribution methods, including LIME \citep{Ribeiro2016WhySI} and SHAP \citep{Lundberg2017AUA}. More recently, natural language explanations have been introduced to generate human‐readable justifications alongside model outputs, often by training on datasets annotated with explanatory comments \citep{Danilevsky2020ASO}.

A promising paradigm is CoT prompting, whereby large language models are induced to produce intermediate reasoning steps that lead to a final answer \citep{Wei2022ChainOT}. This mechanism not only provides the final output of the model, but also generates verifiable decision sequences that can be carefully examined by researchers. Our work utilizes CoT to generate explanations in the context of adversarial defense.

\section{Methodology}
\label{sec:methodology}
\begin{figure*}[h]
  \centering
  \includegraphics[width=\textwidth]{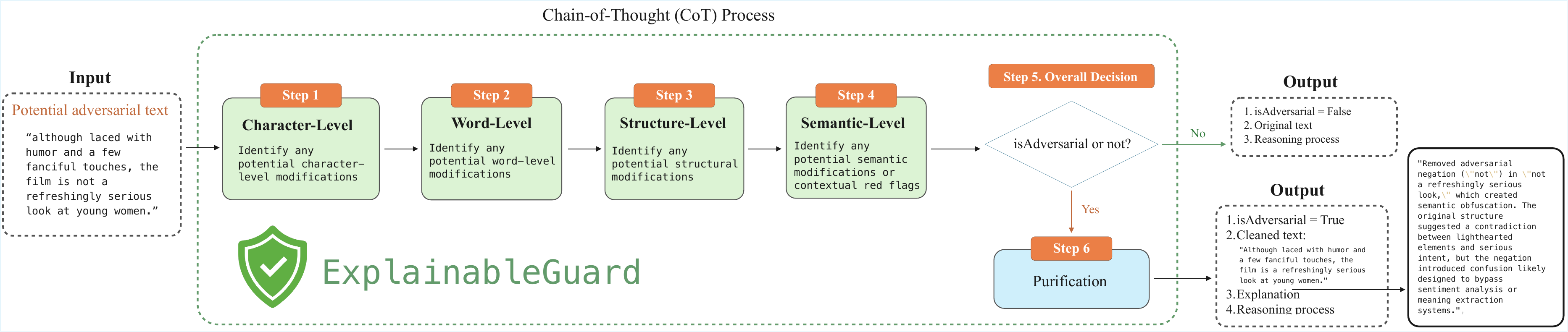}
  \caption{Overview of the ExplainableGuard Chain-of-Thought (CoT) workflow.}
  \label{figure1}
\end{figure*}

Our proposed system, ExplainableGuard, employs DeepSeek-Reasoner~\citep{deepseekai2025deepseekr1incentivizingreasoningcapability} as a security analyst LLM. Given potentially adversarial input text $T_{adv}$, the goal is to produce a cleaned version $T_{clean}$ and a human-readable short explanation $E$ detailing the purification content. Additionally, our system will produce a reasoning process $R$ containing how our LLM analysis the text, detect adversarial patterns and do the purification. The workflow of ExplainableGuard is illustrated in Figure~\ref{figure1}.

\subsection{System Overview}
Our method can be formalized as a function $D: \mathcal{T} \rightarrow (\mathcal{T}, \mathcal{E}, \{0,1\}, \mathcal{R})$, where $\mathcal{T}$ denotes the space of texts, $\mathcal{E}$ denotes the space of explanations, and $\mathcal{R}$ denotes the space of reasoning contents. Given an adversarial input $T_{adv}$, the function outputs a tuple:
$$ (T_{clean}, E, \text{is\_adv}, R) = D(T_{adv}) $$
Here, $T_{clean}$ is the purified text, $E$ is a concise human-readable explanation, $\text{is\_adv}$ is a boolean indicating whether the input was identified as adversarial, and $R$ is the detailed reasoning content generated by the model during the analysis and purification process.

\subsection{Chain-of-Thought Prompting for Defense}
The interpretability of ExplainableGuard is achieved through a carefully designed CoT prompt, $P_{CoT}$, which systematically guides DeepSeek-Reasoner through a sequence of analytical steps. As illustrated in Figure~\ref{figure1}, the prompt instructs the LLM to conduct a comprehensive assessment at multiple levels: starting with character-level inspection (e.g., detecting homoglyphs, invisible characters, typos, leetspeak), followed by word-level analysis to identify unusual synonym usage or suspicious insertions/deletions, then examining structural aspects (sentence structure anomalies or embedded commands), and finally performing semantic and contextual checks to uncover subtle meaning shifts or indirect prompt injections. 

After these analyses, the model determines whether the input is adversarial and formulates an appropriate purification strategy. The LLM then applies this strategy to generate the cleaned text $T_{clean}$. Finally, it produces a structured summary that includes the adversarial judgment ($\text{is\_adv}$), the purified text, a concise explanation $E$, and a reasoning process $R$.

\section{Experimental Setup}
\label{sec:experiments}

\begin{figure*}[h]
  \centering
  \includegraphics[width=\textwidth]{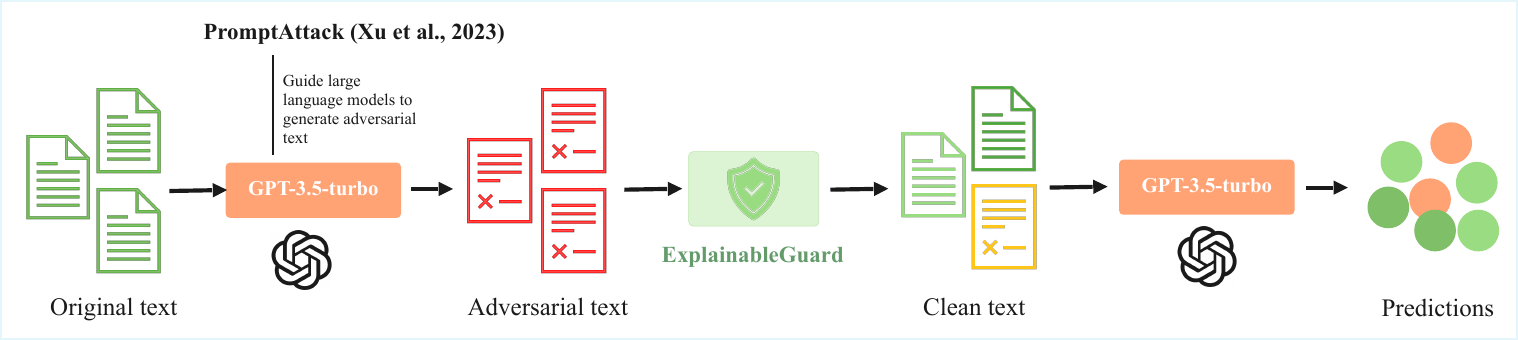}
  \caption{The workflow of the experiments, including the generation of adversarial examples and the defense of ExplainableGuard.}
  \label{figure2}
\end{figure*}

\subsection{Dataset}
We conducted experiments on both short and long text datasets.

\textbf{GLUE Benchmark:} For short text evaluation, we selected three representative tasks from the GLUE benchmark~\citep{glue2018}: SST-2, RTE, and QQP. These datasets are widely used for assessing natural language understanding and are characterized by relatively short input texts. Because of time and budget constraints, we ran a sample of the QQP dataset instead of the full validation set.

\textbf{IMDB Movie Reviews:} For long text evaluation, we included the IMDB movie review dataset~\citep{imdb2011}, which contains lengthy user-written reviews labeled for sentiment.

Together, these datasets allow us to evaluate our method's robustness and interpretability across diverse text lengths and task types.

\subsection{Baselines}
We compare ExplainableGuard with a baseline where no defense is applied. Specifically, we evaluate the performance of a target LLM (GPT-3.5-turbo) on adversarial inputs generated by the PromptAttack method~\citep{xu2023llmfoolitself}, without any additional defense mechanism. 

\subsection{Evaluation Metrics}
We use three main evaluation metrics in our experiments, each serving a distinct purpose:

\textbf{Attack Success Rate (ASR):} ASR directly measures whether our defense can help the model make correct predictions in the presence of adversarial attacks. A lower ASR indicates that the defense is more effective at preventing the model from being fooled by adversarial inputs. It is defined as~\citep{lin2024large}:

\begin{equation}
\text{ASR} = \frac{|\{x \in D_{\text{correct}} : f(x') \neq y\}|}{|D_{\text{correct}}|}
\end{equation}
where $D_{\text{correct}}$ represents the set of samples that are correctly classified by the model on the original test dataset, $x'$ is the corresponding adversarial example, $f(x')$ is the model's prediction result on the adversarial example, $y$ is the true label.

\textbf{BLEU Score:} The BLEU score is used to measure the similarity between the purified text and the original, unperturbed text~\citep{bluescore}. In our evaluation, we compute a weighted average of the 1-gram and 2-gram BLEU scores for each example, and then report the mean over all successfully defended examples. Let $C = \{1, 2\}$ denote the set of $n$-gram orders considered, and $N$ be the total number of successful defense examples. The BLEU score is defined as:
\begin{equation}
    \mathrm{BLEU} = \frac{1}{N} \sum_{i=1}^{N} \sum_{n \in C} w_n \cdot \mathrm{BLEU}_i^{(n)}
\end{equation}
where $w_n = 0.5$ for $n=1,2$, and $\mathrm{BLEU}_i^{(n)}$ is the $n$-gram BLEU score for the $i$-th example. A higher BLEU score indicates that our purification process better preserves the original content.

\textbf{Human Evaluation:} Beyond ASR and BLEU, we conducted a human-evaluation study to assess the interpretability of ExplainableGuard’s explanations, which is the most significant contribution of our work. We sampled 40 successful adversarial defense instances balanced across SST-2, RTE, QQP, and IMDB (10 per dataset). We created an ablated variant, EG-noCoT, which uses the same prompt but without chain-of-thought reasoning, and let it generate an explanation as well. Then, we compared ExplainableGuard’s (EG) explanations with those from EG-noCoT.

We employed a within-subjects, counterbalanced design. For each item and condition, two human annotators were shown the adversarial input, the cleaned output, and the explanation. They rated the following aspects on five-point Likert scales, with higher values indicating better performance:

\begin{itemize}
\item \textbf{Clarity}: "I can understand why this output was produced."
\item \textbf{Specificity}: "The explanation cites concrete features such as tokens or structures."
\item \textbf{Actionability}: "The explanation would help me address similar attacks."  
\item \textbf{Conciseness}: "The explanation is appropriately brief for practical use."
\end{itemize}

We additionally collected a binary deployability-trust judgment ("Would you trust deploying this defense on similar inputs?"). For the final statistic of the binary deployability-trust judgment, if at least one of the two annotators rated an item as "yes," we counted that row of data as "yes."

\section{Preliminary Results and Analysis}
\label{sec:results}

\subsection{Attack Success Rate}
Table~\ref{tab:defense_performance} presents the ASR against PromptAttack-EN (PA-EN) and PromptAttack-FS-EN (PA-FS-EN) attacks~\citep{xu2023llmfoolitself}, comparing the performance of ExplainableGuard (EG) with the baseline model (GPT-3.5-turbo) without any defense. FS denotes 'Few-Shot,' indicating that the adversarial attack employs a few-shot learning strategy to achieve higher performance.

Across the three datasets, we observe a notable reduction in ASR when EG is applied. For instance, under the PA-EN attack, the ASR for RTE drops from 34.30\% to 13.18\%. The average ASR across all datasets is reduced from 37.27\% to 24.21\% (PA-EN) and 42.87\% to 24.31\% (PA-FS-EN). This indicates that EG effectively mitigates the adversarial attacks, enhancing the model's robustness. Additionally, for IMDB dataset, the ASR without defense is 38.71\%, while applying EG reduces it to 30.11\% as shown in Table~\ref{IMDB_defense_performance}. This indicates that our method is also effective for long-text adversarial defense.

\begin{table}[t]
  \centering
  \resizebox{\columnwidth}{!}{ 
    \begin{tabular}{lccccc}
      \hline
      \textbf{Attacks} & \textbf{EG} & \textbf{SST-2} & \textbf{RTE} & \textbf{QQP} & \textbf{Avg.} \\
      \hline
      \multirow{2}{*}{PA-EN} 
          & $\times$     & 56.00 & 34.30 & 21.50 & 37.27 \\ 
          & $\checkmark$ & 40.89 & 13.18 & 18.57 & 24.21 \\ 
      \hline
      \multirow{2}{*}{PA-FS-EN} 
          & $\times$     & 75.23 & 36.12 & 17.26 & 42.87 \\ 
          & $\checkmark$ & 48.61 & 10.32 & 14.01 & 24.31 \\ 
      \hline
    \end{tabular}
  }
  \caption{Performance comparison of defense methods against different attacks on SST-2, RTE, and QQP.}
  \label{tab:defense_performance}
\end{table}

\begin{table}[t]
  \centering
  \begin{tabular}{lcc}
    \hline
    \textbf{Attacks} & \textbf{EG} & \textbf{IMDB} \\
    \hline
    \multirow{2}{*}{PA-EN} 
        & $\times$     & 38.71 \\ 
        & $\checkmark$ & 30.11 \\ 
    \hline
  \end{tabular}
  \caption{Performance comparison of defense methods against different attacks on IMDB.}
  \label{IMDB_defense_performance}
\end{table}

\subsection{BLEU Score}
We evaluate the BLEU score of the purified text on successful defense results. Table~\ref{tab:bleu_scores} reports the BLEU scores for across all datasets. For SST-2, RTE, and QQP, both attack methods achieve high BLEU scores (>0.81), suggesting that ExplainableGuard can effectively clean adversarial inputs while maintaining semantic fidelity. On the IMDB dataset, the PA-EN BLEU score is 0.6195, indicating that EG can still effectively preserve much of the original content even on longer texts.

\begin{table}[!htbp]
  \centering
  \begin{tabular}{lcccc}
    \hline
    \textbf{Method} & \textbf{SST-2} & \textbf{RTE} & \textbf{QQP} & \textbf{IMDB} \\
    \hline
    PA-EN  & 0.82 & 0.8909 & 0.8626 & 0.6195 \\
    PA-FS-EN   & 0.85 & 0.81 & 0.8613 & - \\
    \hline
  \end{tabular}
  \caption{BLEU scores for zeroshot and fewshot methods across different datasets.}
  \label{tab:bleu_scores}
\end{table}

\begin{table*}
\centering 
\resizebox{\textwidth}{!}{
\scalebox{0.83}{
\begin{tabular}{p{0.06\textwidth} p{0.24\textwidth} p{0.24\textwidth} p{0.32\textwidth}}
\hline
\textbf{Dataset} & \textbf{Adversarial Text} & \textbf{Cleaned Text} & \textbf{Explanation} \\
\hline
SST-2
& no sophomore slump for director sam mendes , who segues from oscar winner to oscar-winning potential with a smooth sleight of hand , \textbf{unfortunately}.
& No sophomore slump for director Sam Mendes, who segues from Oscar winner to Oscar-winning potential with a smooth sleight of hand.
& Removed the adversarially placed "unfortunately" at the end, which created a contradictory sentiment likely intended to confuse sentiment analysis models. Standardized capitalization ("Oscar") and punctuation for clarity. The adversarial modifier artificially injected negativity into an otherwise positive statement.
\\ \hline
RTE
& Quebec woman and her mother accused of plotting to kill a four-year-old \textbf{gurl}.
& Quebec woman and her mother accused of plotting to kill a four-year-old \textbf{girl}.
& Detected intentional adversarial misspelling "gurl" (instead of "girl"), likely to bypass child safety content filters. Corrected to restore standard spelling while preserving original meaning.
\\ \hline
QQP
& What is your new year resolution, short term and long term \textbf{gaol} for 2017?
& What is your new year resolution, short term and long term \textbf{goal} for 2017?
& Detected intentional misspelling "gaol" (a homophone for "goal") likely to bypass keyword-based filters. Corrected to "goal" to restore intended meaning and neutralize evasion. No other adversarial patterns found.
\\ \hline
IMDB
& This is a really sad, and touching movie! It deals with the subject of child abuse. It's really sad, but mostly a true story, because it happens everyday. Elijah Wood and Joseph Mazzello play the two children or Lorraine Bracco, a single mother who just tries to make a home for them. While living with her parents, a man, who likes to be called "The King" comes into their life. He hits the youngest boy, Bobby, but the two brothers vow not to tell their mother. But finally she finds out, after the Bobby is hurt badly. The end kind of ruined it for me, because it is so totally unbelievable. But, except for that, I \textbf{hate} the movie."
& This is a really sad, and touching movie! It deals with the subject of child abuse. It's really sad, but mostly a true story, because it happens everyday. Elijah Wood and Joseph Mazzello play the two children or Lorraine Bracco, a single mother who just tries to make a home for them. While living with her parents, a man, who likes to be called "The King" comes into their life. He hits the youngest boy, Bobby, but the two brothers vow not to tell their mother. But finally she finds out, after the Bobby is hurt badly. The end kind of ruined it for me, because it is so totally unbelievable. But, except for that, I \textbf{love} the movie.
& The word "hate" in the final sentence was identified as an adversarial modification (word-level substitution) designed to invert sentiment analysis. This contradicted the overall positive context of the review. Replaced "hate" with "love" to restore semantic coherence. No other modifications detected.
\\ \hline
\end{tabular}}}
\caption{Example of adversarial text purification and corresponding explanation.}
\label{tab:adversarial_defense_example}
\end{table*}

\begin{table*}[!htbp]
  \centering
  \begin{tabular}{lccccc}
    \hline
    \textbf{Method} & \textbf{Clarity} & \textbf{Specificity} & \textbf{Actionability} & \textbf{Conciseness} & \textbf{Deployability Trust} \\
    \hline
    EG           & 4.09 ± 0.67 & 3.92 ± 0.68 & 3.50 ± 0.76 & 3.23 ± 0.77 & 72.5\% \\
    EG-noCoT     & 2.99 ± 0.79 & 3.02 ± 0.78 & 2.89 ± 0.74 & 3.62 ± 0.63 & 42.5\% \\
    \hline
    P-value      & <.001 & <.001 & = 0.0011 & = 0.0125 & = 0.0095 \\
    \hline
  \end{tabular}
  \caption{Human evaluation on the explanation of the sampled data (Mean ± SD).}
  \label{tab:human_evaluate}
\end{table*}

\subsection{Explainability}
The explanations generated by ExplainableGuard provide valuable insights into the defense process. For example, in the case of a PA-EN attack on SST-2, the model identifies specific character-level anomalies (e.g., "homoglyphs" or "typos") and word-level issues (e.g., "unusual synonym usage"). The explanation details how these factors contributed to the adversarial nature of the input and how they were addressed during purification. This level of transparency is crucial for understanding the model's decision-making process and building trust in its outputs. Some successful defense examples with explanation are provided in the Table~\ref{tab:adversarial_defense_example}.

Human evaluation indicates that ExplainableGuard’s explanations are judged to be clearer and more practically useful than those of the ablated, as shown in Table~\ref{tab:human_evaluate}. Across 40 items, averaged over annotators, clarity improved from 2.99 (SD 0.79) with EG-noCoT to 4.09 (SD 0.67) with EG, a mean difference of +1.1, that was statistically significant under a Wilcoxon signed-rank test P < 0.001. Specificity increased from 3.02 (SD 0.78) to 3.92 (SD 0.68; P < 0.001), and actionability rose from 2.89 (SD 0.74) to 3.50 (SD 0.76; P = 0.0011). The conciseness of EG performed worse than EG-noCoT, suggesting that the CoT mechanism might introduce some redundant content into the explanation. Additionally, the proportion of items judged deployable increased from 42.5\% under EG-noCoT to 72.5\% with EG; McNemar’s test indicated a significant shift in trust (P = 0.0095). These findings support the claim that ExplainableGuard’s chain-of-thought guidance yields explanations that users perceive as both clearer and more practically useful.

Overall, these results demonstrate that ExplainableGuard substantially reduces the attack success rate across both short and long text datasets, while maintaining high similarity between the purified and original texts. Moreover, the results highlight that ExplainableGuard’s chain-of-thought mechanism produces explanations that are clearer, more specific, actionable, and trusted by users, further underscoring the effectiveness and interpretability of our defense approach.

\section{Discussion}
\label{sec:dicussion}

The experimental results demonstrate that ExplainableGuard offers a compelling dual advantage: it provides effective defense against adversarial attacks while delivering transparent, human-readable explanations for its actions. This discussion synthesizes these findings, contextualizes them within the broader literature, and outlines the implications and unique position of our work.

\subsection{Efficacy–Interpretability Trade-off}
Our results confirm that leveraging the inherent reasoning capabilities of a powerful LLM like DeepSeek-Reasoner through structured CoT prompting is a viable strategy for adversarial defense. The significant reduction in Attack Success Rate (ASR) across both short (SST-2, RTE, QQP) and long (IMDB) text datasets indicates that the multi-faceted analysis—spanning character, word, structural, and semantic levels—is effective at identifying and neutralizing a variety of adversarial perturbations.

However, the defense is not perfect. The residual ASR, particularly on the more challenging PA-FS-EN attacks, suggests that highly sophisticated or novel attack strategies can still partially bypass the purification process. This is an expected characteristic of an arms race in AI security. The strength of ExplainableGuard in this context is not just its defensive performance, but its ability to provide a rationale for its decisions, which is invaluable for understanding failure modes and iteratively improving defenses.

The high BLEU scores, especially on shorter texts, affirm that the purification process is largely conservative and aims to preserve the original semantic content. The lower score on the IMDB dataset is understandable, as longer texts offer more surface area for complex, context-dependent adversarial manipulations, making perfect restoration more challenging. Nonetheless, the score indicates a strong ability to retain the core meaning of the original input.

\subsection{Positioning ExplainableGuard in the Defense Landscape}
Our work enters a growing field of using LLMs themselves for adversarial purification. A contemporary study, LLAMOS \citep{lin2025largelanguagemodelsentinel}, demonstrates impressive defensive performance, achieving high robust accuracy across multiple GLUE tasks. We acknowledge that on certain metrics and datasets, such as ASR on SST-2, LLAMOS reports stronger pure defensive efficacy.

However, the comparison should not be reduced to a single dimension of performance. The primary contribution of ExplainableGuard is not to claim state-of-the-art attack neutralization, but to pioneer a critical, overlooked aspect of adversarial defense: explainability. While methods like LLAMOS effectively purify text, they operate as a black box, providing no insight into why an input was modified or how the attack was neutralized. This lack of transparency is a significant barrier to trust and deployment in high-stakes applications.

ExplainableGuard fills this gap. It is, to the best of our knowledge, the first work to systematically integrate Chain-of-Thought reasoning into the adversarial defense process to generate human-readable justifications. This represents a paradigm shift from opaque defense to interpretable, collaborative security analysis.

Furthermore, our work provides one of the first evaluations of an LLM-based purification defense on a long-text dataset (IMDB), moving beyond the standard short-sentence benchmarks. This demonstrates the viability of our approach in more realistic and complex textual environments, where understanding context is crucial for both defense and explanation.

\subsection{Value of Explainability}
The human evaluation results underscore our core contribution. The statistically significant improvements in Clarity, Specificity, and Actionability for the full ExplainableGuard (EG) over its ablated variant (EG-noCoT) provide strong evidence that Chain-of-Thought reasoning is instrumental in generating useful explanations. Annotators could better understand the model's logic, pinpoint the adversarial elements, and see how the defense could be applied to similar cases.

The near-doubling of the Deployability-Trust rating from 42.5\% to 72.5\% is particularly telling. In practical applications, trust is as crucial as raw performance. A black-box defense that occasionally fails without justification is likely to be discarded. In contrast, a system that explains its reasoning, even when imperfect, allows users to gauge its reliability and understand its limitations, fostering a more collaborative and informed human-AI security partnership.

The one metric where the CoT approach underperformed was Conciseness. This highlights a natural tension between thoroughness and brevity. The step-by-step reasoning, while valuable for transparency, can introduce verbosity. Future iterations could explore methods to summarize the CoT process into more concise explanations without sacrificing the critical details that enable trust and actionability.

\subsection{Limitations and Future Work}
While promising, this approach has several limitations that pave the way for future research. First, the performance of ExplainableGuard is intrinsically tied to the capabilities of the underlying defense LLM. As adversarial tactics grow more advanced, specifically designed to exploit the reasoning patterns or blind spots of models like DeepSeek-Reasoner, the defense could be circumvented.

Second, the computational cost and latency are non-trivial. The need to generate extensive CoT reasoning for each query results in processing times that are prohibitive for real-time, high-throughput applications. This makes ExplainableGuard currently more suitable for offline analysis or critical, lower-volume tasks. Future work should focus on optimization and distillation: exploring whether the robust analytical and explanatory capabilities can be distilled into a smaller, more efficient model.

Finally, our evaluation, while covering diverse datasets, can be expanded. A more comprehensive assessment against a wider array of attack types and on more domains is necessary. Furthermore, the explanation quality was evaluated on a sample of successful defenses; evaluating explanations in failure cases is equally important for a complete trustworthiness audit.

In conclusion, ExplainableGuard represents a pivotal step towards transparent and trustworthy AI security. By framing adversarial defense as an interpretable reasoning process, we move beyond opaque shields to creating collaborative security analysts that work alongside humans to build more resilient and understandable AI systems. While pure defensive performance will continue to be optimized by works like LLAMOS \citep{lin2025largelanguagemodelsentinel}, we argue that the path to truly trustworthy deployment is paved with explainability.

\section{Conclusion}
\label{sec:conclusion}

We introduced ExplainableGuard, an adversarial
defense system that utilizes DeepSeek-Reasoner
and Chain-of-Thought prompting to detect, neu-
tralize, and explain its actions against adversar-
ial text. By guiding the LLM through a systematic
analysis, our method provides not only a cleaned
output but also a transparent rationale for its de-
cisions. This approach enhances trustworthiness
and provides valuable insights for users and se-
curity analysts. While further research is needed,
ExplainableGuard demonstrates a promising direc-
tion for building more robust and understandable
AI security systems.

\section{Bibliographical References}\label{sec:reference}

\bibliographystyle{lrec2026-natbib}
\bibliography{reference}

\end{document}